# CFD Modelling and Sensitivity-Guided Design of Silicon Filament CVD Reactors

G.P. Gakis[1,*], G. Loachamín-Suntaxi[1,2], I.G. Aviziotis[1,4], E.D. Koronaki[3],

P.P. Filippatos[4], G. Chatzigiannakis[4],

S. P. A. Bordas,[2], D. Davazoglou[4], A.G. Boudouvis[1]

[1] School of Chemical Engineering, National Technical University of Athens (NTUA), Heroon Polytechniou 9, 15780 Zografou, Greece

[2] Faculty of Science, Technology and Medicine, University of Luxembourg, Esch-sur-Alzette, L-4364, Luxembourg

[3] Luxembourg Institute of Science and Technology (LIST), Esch-sur-Alzette, L4364, Luxembourg

[4] NCSR "Demokritos", Institute of Nanoscience and Nanotechnology, POB 60228, 153 10 Agia Paraskevi Attiki, Greece.

*e-mail: gakisg@chemeng.ntua.gr

**Abstract**

Filament-based chemical vapor deposition (CVD) for silicon (Si) coatings is often treated as a straightforward adaptation of planar deposition—just with a cylindrical substrate. But this overlooks a fundamental shift in how transport phenomena and reaction kinetics interact. In filament CVD, the filament is not just a substrate; it's the dominant heat source and flow disruptor all in one. In this work, we ask: *What really governs Si film growth on filaments?* Using a validated three-dimensional computational fluid dynamics (CFD) model, we show that filament geometry, thermal gradients, and flow-induced buoyancy do not merely affect uniformity, they define the very regimes (reaction-limited, transition, diffusion-limited) in which deposition occurs. Our model, validated against three independent experimental studies ($R^2 = 0.969$), reveals how temperature, flow rate, and reactor layout fundamentally shift the growth dynamics. We show that reducing filament diameter from 2 mm to 500 μm doesn't just increase local surface area—it can triple growth rates (up to ~39 μm/min), but only in carefully tuned high-temperature regimes. Similarly, multi-filament setups don't simply scale deposition—they reshape flow fields and compound thermal asymmetries, sometimes improving uniformity, other times worsening it. These effects are not evident without modeling the interplay between transport and kinetics, something experiments alone can't resolve. To bridge physical understanding with design actionability, we apply global sensitivity analysis via Polynomial Chaos Expansion (PCE) and Sobol' indices. These reveal how control shifts: in reaction-limited regimes, temperature dominates; in diffusion-limited regimes, reactant transport takes over. By making these interdependencies visible and quantifiable, our model lays the foundation for design strategies that are evidence-based, not assumption-driven.

**Keywords**: CVD reactor; CFD model; sensitivity analysis; filament deposition; computer-aided design

## 1. Introduction

Silicon (Si) thin films are foundational to technologies ranging from microelectronics and energy storage to photovoltaics and sensor platforms [1–4]. This technological breadth has driven persistent interest in scalable methods for producing Si-based materials—whether as thin films [7], nanoparticles [8], nanotubes [9], nanowires [10], or continuous fibers [11]. Among these, Chemical Vapor Deposition (CVD) remains the dominant technique for high-purity Si film growth, particularly in epitaxial processes using chlorosilane (e.g., $SiCl_4$, $SiH_xCl_y$) or silane ($SiH_4$) precursors in hydrogen-rich environments [12–17].

While planar substrates have long been the standard in both industrial and academic contexts, growing demand for conformal deposition on non-planar structures—such as wires and fibers—has shifted attention toward filament-based CVD [18–21]. In this configuration, the filament itself acts as both the substrate and the thermal driver, resistively heated to activate precursor decomposition in its immediate vicinity. This design offers compelling advantages: localized heating, reduced thermal budget, and potentially higher growth rates. However, these benefits come at the cost of introducing fundamentally different deposition dynamics.

The CVD of filament-based systems exhibit steep thermal gradients, buoyancy-driven flow recirculation, and spatially varying precursor concentrations. The filament acts not as a passive surface but as an active agent that defines the local flow-thermal-chemical environment. These coupled effects produce growth regimes that diverge sharply from the classical planar case—and invalidate many of its core assumptions. Nonetheless, most existing studies fail to capture this complexity, often decoupling chemistry from physics or relying on 2D simplifications that ignore the filament's role as a spatial and thermal driver.

In this context, computational modelling has emerged as a valuable tool to investigate such processes [22-25]. Using such approaches, the interplay between the different mechanisms that constitute the CVD process can be investigated in detail [26-28], while the key aspects and the limiting mechanisms of the process can be identified. For the case of Si thin films, some works have tried to perform such computational studies, from atomistic scale models[29,30], to thermodynamic[31, 32], kinetic[17, 33, 34], and reactor scale studies[12, 35], with multiscale studies also being the focus of some works[36, 37]. However, the developed models focus mainly on the chemical reaction mechanisms and their kinetics [17], while their interplay with transport phenomena is often overlooked. Furthermore, the models developed for Si films consider the case of planar substrates only, while the deposition on the surface of filaments has not been studied computationally.

To bridge this gap, we develop a fully coupled, three-dimensional Computational Fluid Dynamics (CFD) model of Si deposition on heated tungsten (W) filaments. The model integrates gas-phase flow, heat transfer, species transport, gas-phase reaction kinetics, and surface deposition within a unified framework. Critically, it is validated against three independent experimental datasets [12, 13, 15], covering different precursors and reactor geometries, which confirms its robustness and applicability.

Beyond predictive accuracy, the model serves as a diagnostic platform to reveal the mechanistic interplay between reaction kinetics and transport phenomena [38-40]. It captures the emergence of non-uniform deposition due to temperature-driven convection and maps the transitions between reaction-limited and diffusion-limited regimes [41]. These spatial patterns and regime boundaries are difficult to infer from experimental averages, but they are central to process optimization and scale-up.

To translate this mechanistic insight into actionable design guidance, we perform a global sensitivity analysis using Polynomial Chaos Expansion (PCE) and Sobol' indices [26]. This analysis quantifies how control parameters—such as filament temperature, hydrogen inflow rate, and precursor concentration—govern variability in deposition rate under different operating regimes. The results offer clear prioritization maps for process tuning depending on the dominant deposition constraints.

In sum, this work reframes filament CVD not as a minor extension of planar systems, but as a distinct reactor architecture that demands a geometry-aware, physically coupled design methodology. Our model establishes a foundation for rigorous, mechanistically grounded optimization and paves the way for knowledge-driven scale-up of filament-based Si deposition to industrially relevant throughputs.

## 2. Methods

### 2.1. Computational domain

The model geometry represents a simplified baseline configuration used as a starting point for exploratory analysis of filament CVD behavior and for fine-tuning the model through comparison with experimental data. It consists of a 100 mm diameter cylindrical chamber with a width of 160 mm. Inlet and outlet tubes, 8 mm and 20 mm in diameter respectively, are connected at either side of the chamber. A single filament is positioned along the chamber's central axis, oriented perpendicular to the gas flow, and supported by two electrodes. The filament length is set to 80 mm, with an initial diameter of 2 mm. This geometry serves both as a foundational test case for investigating the influence of reactor layout and process parameters, and as a calibration framework to validate the physical fidelity of the CFD model. A schematic of the chamber is shown in Figure 1.

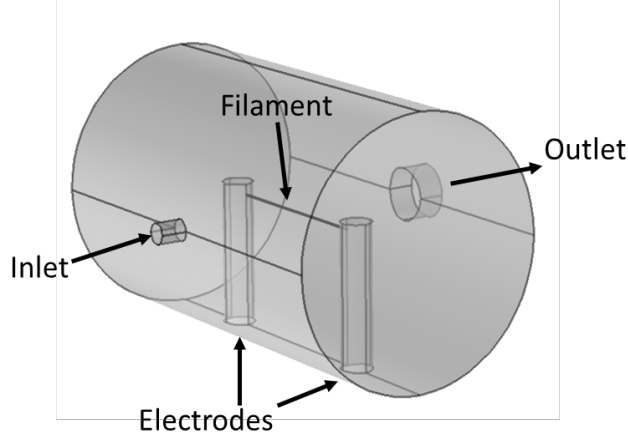

Figure 1. Schematic of the filament CVD reactor geometry.

## 2.2. Governing equations

The gas mixture flow inside the filament CVD reactor is modeled using the continuum medium hypothesis. This hypothesis is validated by calculating the Knudsen number [42,43], defined as:

$$Kn = \lambda/L \qquad (1)$$

$$\lambda = \frac{k_B T}{\sqrt{2}\pi d^2 P}, \qquad (2)$$

where λ is the mean free path of the gas particle and L is the characteristic length, taken equal to the reactor diameter in the present case. $k_B$ is the Boltzmann constant, T is the gas-phase temperature, d is the gas species particle diameter, and P is the process pressure. For the present study, the continuum medium hypothesis is valid since the maximum $Kn_{max}$= $5.6 \cdot 10^{-6}$ <<0.1 [44].

The flow inside the filament CVD reactor tube can be assumed to be laminar, as indicated by the Reynolds number for fluid flow within a pipe[24, 28]:

$$Re = \frac{\rho \cdot u \cdot D}{\mu}, \qquad (3)$$

where ρ is the mixture density, u is the gas velocity, D is the reactor diameter, and μ is the dynamic viscosity of the gas. In the present study, $Re_{max}$=150. The low value calculated for the Reynolds number validates the laminar flow assumption. The gas mixture is assumed to behave like an ideal gas.

Based on these assumptions, the governing equations for the description of transport phenomena inside the filament CVD reactor include the conservation of mass, momentum, and energy, coupled with the conservation of chemical species, which are usually applied for the modeling of deposition processes [43, 45].

**Conservation of mass**

$$\frac{\partial \rho}{\partial t} = \nabla \cdot (\rho u) = 0 \tag{4}$$

**Conservation of momentum**

$$\frac{\partial (\rho u)}{\partial t} + \nabla \cdot (\rho u u) = -\nabla P + \nabla \cdot \left[ \mu (\nabla u + \nabla u^T) - \mu \frac{2}{3} (\nabla \cdot u) I \right] + \rho g \tag{5}$$

where **u** is the velocity vector, **I** is the unit tensor and **g** the gravity acceleration

**Conservation of energy**

$$C_p \frac{\partial (\rho T)}{\partial t} + C_p \nabla \cdot (\rho u T) = \nabla \cdot (k \nabla T) \tag{6}$$

where $C_p$ is the specific heat of the gas mixture and k is the thermal conductivity.

**Conservation of chemical species**

$$\frac{\partial (\rho \omega_i)}{\partial t} + \nabla \cdot (\rho u \omega_i) = -\nabla \cdot j_i + R_i \tag{7}$$

where $\omega_i$ is the mass fraction of the i species in the gas phase.

The diffusion flux $j_i$ is calculated:

$$j_i = -\rho \omega_i \sum_{k=1}^{n-1} D_{ik} \left[ \nabla x_k + (x_k - \omega_k) \frac{\nabla P}{P} \right] - D_{T,i} \frac{\nabla T}{T} \tag{8}$$

The thermal conductivity, dynamic viscosity, and diffusion coefficients for the chemical species are calculated using the kinetic gas theory [22, 46]. The Lennard-Jones parameters for the species are obtained from the CHEMKIN-PRO database [47]:

$$\mu_i = 2.669 \cdot 10^{-6} \frac{\sqrt{T M_i \cdot 10^{-3}}}{\sigma_i \Omega_D}, \tag{9}$$

where $\mu_i$ is the dynamic viscosity, $\sigma_i$ is the Lennard-Jones characteristic diameter and $\Omega_D$ is the collision integral for viscosity, defined as a function of the dimensionless temperature as $\Omega_D = f(T k_B / \varepsilon_i)$, with $\varepsilon_i / k_B$ being the Lennard-Jones energy potential.

The binary diffusion coefficient can be given by the following expression:

$$D_{i,j} = 2.695 \cdot 10^{-3} \frac{\sqrt{T^3 (M_i + M_j)/(2 \cdot 10^{-3} M_i M_j)}}{\rho \sigma_i \sigma_j} \cdot f\left(\frac{\varepsilon_k}{k_B}\right), \tag{10}$$

with $\varepsilon_k = \sqrt{\varepsilon_i \varepsilon_j / k_B^2}$, and the thermal conductivity is calculated by eq. 7, derived from the kinetic theory of gases:

$$k_i = 2.669 \cdot 10^{-6} \frac{\sqrt{T M_i \cdot 10^{-3}}}{\sigma_i^2} \cdot \frac{1.15 C_{p,i} + 0.88 R}{M_i \Omega_D}, \tag{11}$$

where $C_{p,i}$ is the specific heat capacity of species i. Finally, the heat capacity for each species is given as a polynomial function of temperature, following the NASA format [48]:

$$\frac{C_P}{R} = a_1 + a_2 T + a_3 T^2 + a_4 T^3 + a_5 T^4. \tag{12}$$

## 2.3. Chemical mechanisms

The chemical reactions involved in silicon (Si) CVD using $SiCl_4$ and $H_2$ as precursors have been thoroughly reviewed in the scientific literature [14]. The overall reaction occurring in the CVD chamber can be written as [15]:

$SiCl_4 + 2H_2 \leftrightarrow Si + 4HCl$ (R1)

Multiple reaction mechanisms have been suggested, based on experimental studies [14, 16, 49] and gas-phase analyses [50]. Additionally, ab initio theoretical models have been employed to estimate reaction kinetics and energy barriers for specific elementary steps [29, 30]. These insights have contributed to the development of reduced-order reaction models incorporating chains of coupled gas-phase and surface reactions [17].

Among these studies, Narusawa et al. [51] identified $SiHCl_3$ and $SiCl_2$ as the dominant intermediate species formed after the dissociation of $SiCl_4$, with $SiCl_2$ being the primary contributor to Si deposition. This conclusion is further supported by ab initio calculations [29], which also point to $SiCl_2$ and $SiHCl_3$ as key intermediates.

Based on these findings, the complex chain of gas-phase reactions can be simplified by lumping intermediate steps into an effective two-step mechanism. While this simplification reduces the model's ability to resolve intermediate species concentrations in detail, it enables a more computationally efficient investigation of the coupled transport and kinetic phenomena within the complex three-dimensional reactor geometry (Figure 1). The lumped gas-phase reaction scheme is:

$SiCl_4 + H_2 \rightarrow SiHCl_3 + HCl$ (R2)

$SiHCl_3 \rightarrow SiCl_2 + HCl$ (R3)

Reactions R2 and R3 are modeled using first order Arrhenius expressions:

$R_{R2} = A_{R2} \cdot \exp[-E_{R2}/(RT)] \cdot C_{SiCl4} \cdot C_{H2}$ (13)

$R_{R3} = A_{R3} \cdot \exp[-E_{R3}/(RT)] \cdot C_{SiHCl3}$ (14)

Following the approach of Narusawa et al. [46], $SiCl_2$ is assumed to be the main precursor for Si deposition on the surface. The different steps of $SiCl_2$ and $H_2$ dissociation on the surface [17], are lumped into one mechanism:

$SiCl_{2(g)} + H_{2(g)} + 4\text{<s>} \rightarrow Si_{(b)} + 2HCl_{(g)} + 4\text{<s>}$ (R4)

Reaction R4 simulated using first order Arrhenius expressions, as in the case of R2 and R3, obtaining the final deposition rate in $mol/(m^2 \cdot s)$. It is mentioned that for the kinetics of reaction R4, the activation energy of the most energetically demanding step of the surface reactions from [17] is used as the activation energy of R4 (67 kcal/mol).

The growth rate, in terms of thickness per minute, is computed based on the following approach:

$$GR = \frac{R_{R4} \cdot M_{Si}}{\rho_{Si}} \qquad (15)$$

Where GR is the growth rate, $R_{R4}$ is the rate of R4, $M_{Si}$ is the molar mass of Si (28g/mol) and $\rho_{Si}$ is the Si density, taken as 2328(kg/m$^3$).

**2.4. Model validation**

The modeling framework was validated against experimental data from three independent studies widely cited in the literature. Each study features distinct precursors and reactor configurations, enabling robust assessment across a representative range of conditions. First, the study by Angermeier et al. [12] was used to calibrate the model under conditions involving SiHCl$_3$ as the sole silicon precursor in a horizontal CVD reactor. Because this setup excludes SiCl$_4$, only reaction R3 is active, allowing direct fitting of its pre-exponential factor to match the experimental deposition rates.

Next, the model was applied to replicate the results of Sugawara [15], who used SiCl$_4$ and H$_2$ in a rotating disk CVD system. This enabled calibration of the pre-exponential factor for reaction R2, isolating its contribution to the overall deposition kinetics. Finally, the model was validated against the work of Bloem [13], which includes deposition cases using both SiCl$_4$ and SiHCl$_3$. This third dataset served as an integrated test of the full gas-phase reaction scheme, R2 and R3, under mixed-precursor conditions.

For each case, reactor geometries were reconstructed based on the descriptions provided in the respective publications, and boundary conditions were assigned to match reported operating parameters. This step ensured that model predictions reflected the physical context of each experiment, enabling direct comparison.

Through this tiered validation procedure, the kinetic parameters for the gas-phase reactions R2 and R3 were systematically estimated and the resulting reaction model was then applied to simulate the current filament-based CVD system.

**2.5. Model implementation**

In addition to the governing equations describing fluid flow, heat transfer, and species transport within the filament CVD reactor, appropriate boundary conditions are imposed to replicate experimental conditions.

A no-slip boundary condition is applied to all solid surfaces, including the reactor walls, filament, and electrodes. The reactor walls are assumed adiabatic, with zero flux for all chemical species. At the inlet, a fully developed flow is prescribed, with gas composition and mass flow rate matching experimental values. The inlet temperature is fixed at 20 °C (Dirichlet condition).

At the reactor outlet, the operating pressure is imposed, along with zero-gradient (outflow) conditions for temperature and species. The filament temperature is applied as a Dirichlet condition, and surface reaction R4 is implemented as a species flux on the filament surface. Gas-phase reactions R2 and R3 are included as volumetric source terms throughout the domain.

The complete set of experimental operating conditions and model parameters is summarized in Table 1.

Table 1 List of model parameter values

| *Parameter* | *Value* | *Description* | *Reference* |
|---|---|---|---|
| $P_{op}$ | 1 atm | Operating pressure | Experiment |
| $T_{in}$ | 20°C | Inlet gas temperature | Experiment |
| $T_{fil}$ | 1100°C | Filament temperature | Experiment |
| $F_{N2}$ | 297 sccm | $N_2$ inlet flow rate | Experiment |
| $F_{H2}$ | 33 sccm | $H_2$ inlet flow rate | Experiment |
| $F_{SiCl4}$ | 40 sccm | $SiCl_4$ inlet flow rate | Experiment |
| $A_{R2}$ | 5E9 | Pre exponential factor (R2) | Fitted to [15] |
| $A_{R3}$ | 1E13.5 | Pre exponential factor (R3) | Fitted to [12] |
| $A_{R4}$ | 1E10.51 | Pre exponential factor (R4) | [17] |
| $E_{R2}$ | 55 kcal/mol | Activation Energy (R2) | [45] |
| $E_{R3}$ | 73.7 kcal/mol | Activation Energy (R3) | [17] |
| $E_{R4}$ | 67 kcal/mol | Activation Energy (R4) | [17] |

To reduce computational cost, geometric symmetry was exploited by simulating only half of the CVD chamber, applying a symmetry boundary condition along the midplane.

The governing equations were discretized over a mesh consisting of 136,273 elements, as shown in Figure 2. Simulations were carried out using the finite element method in COMSOL Multiphysics®, with linear basis functions for all variables. The mesh resolution was selected following a mesh independence study to ensure numerical accuracy and solution convergence.

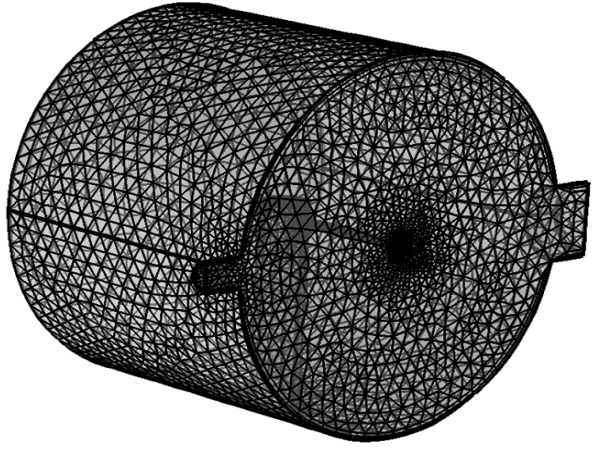

Figure 2. Finite element mesh (136,273 elements) applied to the symmetric half of the reactor geometry, used for all simulations.

**2.6. Sensitivity analysis via Polynomial Chaos Expansion (PCE) and Sobol' indices**

To identify the most influential process parameters and understand their interactions, a global sensitivity analysis was performed using Sobol' indices. These indices decompose the output variance of a model into contributions from individual input parameters and their interactions, offering a robust, quantitative measure of parameter influence.

To efficiently compute the Sobol' indices, we constructed a Polynomial Chaos Expansion (PCE) surrogate model [52, 53] of order dNd_NdN, which approximates the model output M(x) as:

$$M_d^{PCE}(x) = c_0 + \sum_{\alpha \in \mathbb{N}^m, |\alpha| \leq d} c_\alpha \psi_\alpha(x),$$

where $m$ is the number of input parameters and $\{\psi_\alpha\}_{\alpha \in \mathbb{N}^m}$ is an orthogonal polynomial basis. This formulation enables variance-based sensitivity analysis [52, 54], as the variance of the model response can be estimated directly from the PCE coefficients:

$$Var(M_d^{PCE}(x)) = \sum_{|\alpha| \leq d, \alpha \neq 0} c_\alpha^2.$$

Additionally, the total-order Sobol' index $S_i^T$ for each input parameter $i = 1, \ldots, m$ can be also expressed in terms of the PCE coefficients as:

$$S_i^T = \sum_{|\alpha| \leq d, \alpha_i \neq 0} c_\alpha^2 / Var(M_d^{PCE}(x)).$$

This index quantifies the total contribution of each input parameter and its interactions to the output variability. Values of $S_i^T$ close to one indicate strong influence, while values near zero imply a negligible effect.

## 3. Results and Discussion
### 3.1. Model validation

As a critical first step in this study, we rigorously validated the computational framework against experimental data from three independent and methodologically diverse studies [12, 13, 15]. This validation serves two primary purposes: (i) to calibrate key kinetic parameters in the reduced gas-phase reaction scheme, and (ii) to establish the credibility and transferability of the CFD-based modeling approach for simulating Si deposition from both $SiCl_4$ and $SiHCl_3$ precursors.

The first benchmark is the study by Angermeier et al. [12], which examines Si deposition from $SiHCl_3$ and $H_2$ in a horizontal CVD reactor. Because this system excludes $SiCl_4$, it provides a clean case to isolate and fit the kinetics of reaction R3. Experimental conditions include a $SiHCl_3:H_2$ ratio of 15%, an inlet velocity of 0.63 cm/s, and atmospheric pressure, with deposition temperatures systematically varied.

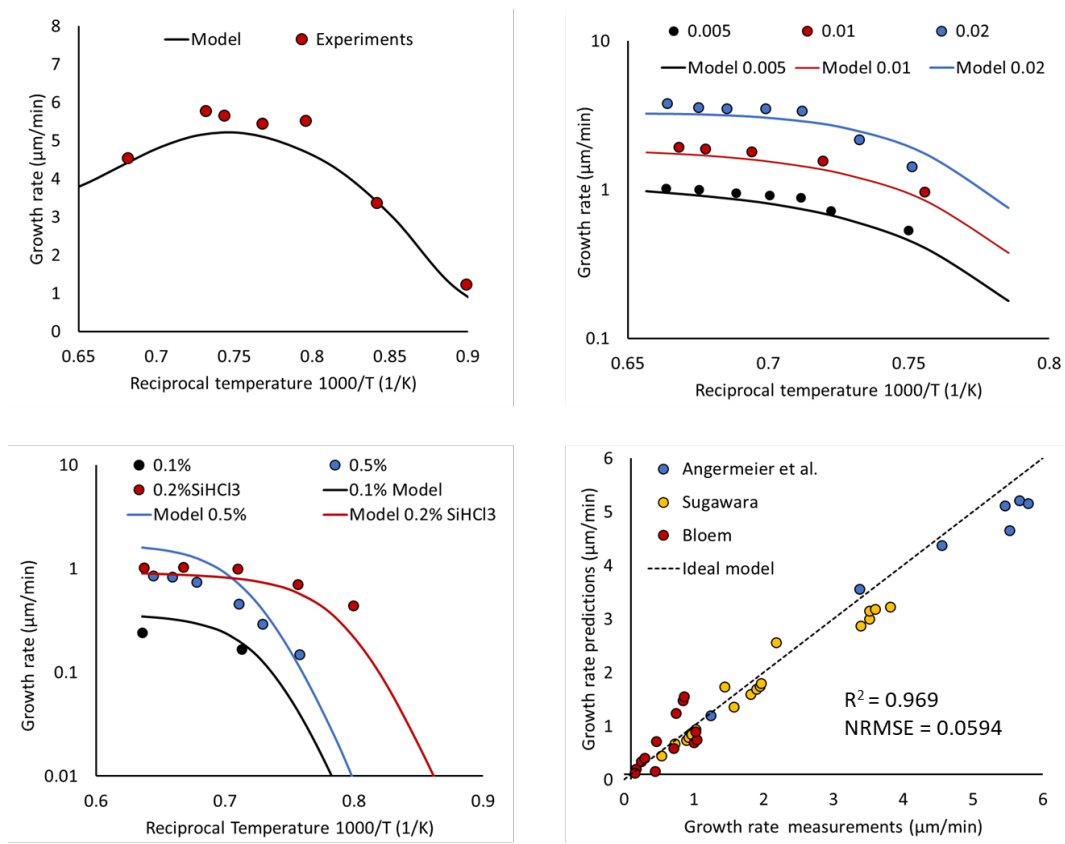

Figure 3. Comparison of model predictions with the experimental measurements of the validation cases of a) Angermeier et al. [12], b) Sugawara [15], c) Bloem [13], d) Overall model fitting

The second validation case is the study by Sugawara [15], which focuses on $SiCl_4$ as the precursor in a rotating disk CVD setup. A substrate rotational speed of 120 rpm and a

40 L/min hydrogen inflow were imposed at atmospheric pressure. In this setup, the $SiCl_4$ mole fraction and temperature were varied. This case enables targeted calibration of reaction R2 and challenges the model to capture the transport-dominated flow dynamics of a rotating substrate. Finally, we tested the model against the study by Bloem [13], which uniquely investigates both $SiCl_4$ and $SiHCl_3$ precursors. For $SiHCl_3$, a 0.2% inlet mole fraction was used; for $SiCl_4$, the inlet fraction was varied between 0.1% and 0.5%, with a constant hydrogen flow of 75 L/min at atmospheric pressure. This dataset evaluates the model's predictive ability when both gas-phase reactions are active and interacting.

In each case, the reactor geometry and boundary conditions were reconstructed to reflect the original experimental setups as closely as possible. The resulting simulations were compared directly to reported deposition rates, providing a rigorous and multi-faceted validation of the model. The agreement between model predictions and experimental data is shown in Figure 3.

The consolidated comparison in Figure 3d illustrates the model's robustness across diverse experimental configurations, precursor chemistries, and flow conditions. A strong correlation is observed between predicted and measured growth rates ($R^2 = 0.969$), and the normalized root-mean-square error remains low (NRMSE = 0.0594), reflecting both the consistency and predictive fidelity of the model.

Minor discrepancies are observed in localized temperature ranges for some datasets—most notably in the high-temperature regime of the Sugawara case (Figure 3a), where the model slightly underpredicts growth rates. This deviation likely stems from unmodeled complexities such as thermal boundary layer thinning near the rotating substrate or localized precursor depletion effects not captured in the reduced reaction mechanism. In the Bloem case (Figure 3c), small deviations are observed at the lowest inlet concentrations, where measurement uncertainty and sensitivity to boundary condition assumptions become more pronounced.

Despite these localized differences, the model consistently captures the dominant physical trends—reaction-limited growth at low temperatures, kinetic saturation, and eventual transition to diffusion-limited behavior. These outcomes, together with the statistical agreement across three independent validation cases, establish a high degree of confidence in using the fitted kinetic model for the subsequent design and analysis of filament-based CVD reactors.

**3.2. Reactor flow dynamics and transport phenomena**

Building on the validated model, we now examine the internal flow structure and transport phenomena that emerge under the operating conditions listed in Table 1. The simulation results (presented in Figure 4) provide spatially resolved insight into the velocity field, temperature distribution, and gas density within the filament CVD reactor.

Figure 4a reveals a nontrivial flow topology driven by the interaction between forced convection and buoyancy. Three distinct recirculation zones are observed: two in the upper half of the reactor on either side of the filament, and one below the filament. The primary inflow enters through the bottom inlet, generating a higher-velocity region in the lower part of the chamber. Meanwhile, the heated filament induces strong thermal gradients, elevating the local gas temperature near its surface (Figure 4b). The reduced density of the heated gas initiates buoyancy-driven ascent, which reinforces upward flow in the central and upper regions of the chamber. This interaction creates stable recirculation zones and a vertically stratified temperature field.

The spatial variation in temperature shown in Figure 4b aligns with this flow behavior: regions near the filament experience elevated temperatures, while cooler zones persist near the chamber base due to the room-temperature inflow. This thermal stratification, driven by buoyancy, leads to a vertically asymmetric thermal environment that is not captured in simplified or planar models.

These dynamics are fundamentally governed by the compressibility of the gas mixture and its temperature-dependent density (Figure 4c). As the local temperature increases, gas density decreases, inducing upward flow due to buoyancy forces. These density gradients are self-reinforcing: heating by the filament alters the local flow, which in turn modifies heat transport and species distribution. This nonlinear coupling between flow, temperature, and density is central to understanding reactor-scale deposition behavior.

Such flow behavior is challenging to isolate experimentally due to the opacity and spatial heterogeneity of the reactor environment. The CFD model reveals that transport phenomena—driven by the geometry-specific interaction of thermal and fluidic forces—can significantly influence species residence time, precursor delivery to the surface, and ultimately the deposition profile.

These insights underscore the importance of accounting for flow symmetry breaking, localized recirculation, and buoyancy effects in filament-based CVD systems. They also highlight the diagnostic value of high-fidelity simulations in exposing transport mechanisms that directly impact growth uniformity and process control. Understanding this interplay is essential for rational reactor design and for optimizing process windows in industrial-scale applications.

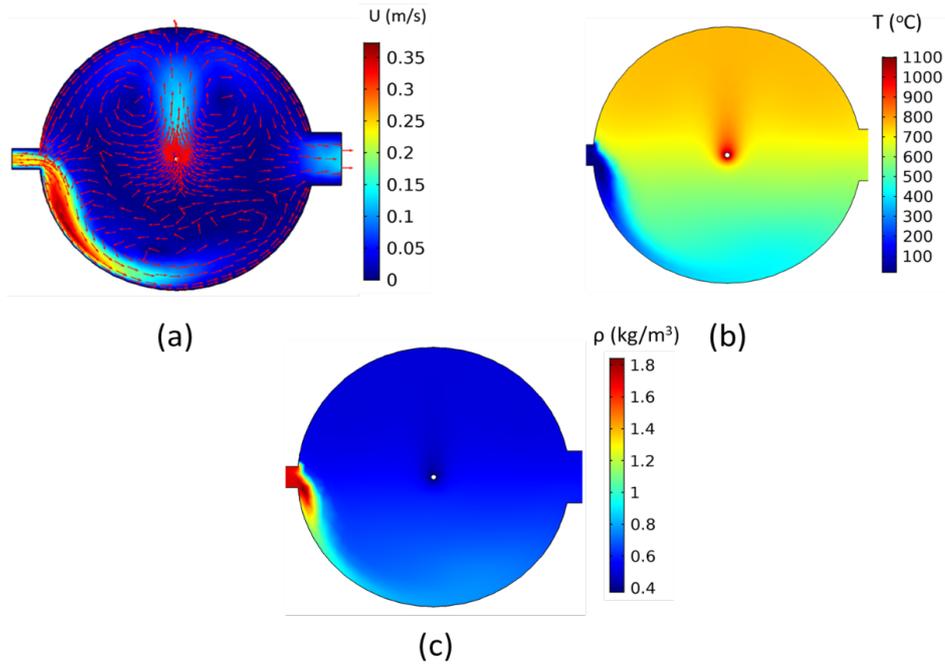

Figure 4. Simulated transport phenomena within the filament CVD reactor under baseline operating conditions. **(a)** Velocity field and streamlines showing the emergence of three distinct recirculation zones: two in the upper region flanking the filament and one below it. The pattern is driven by buoyancy and filament heating, coupled with bottom inflow. **(b)** Temperature distribution illustrating thermal stratification due to localized heating by the filament and cooling from room-temperature inflow. The hottest region lies just above the filament, coinciding with upward convective motion. **(c)** Gas density field, inversely correlated with temperature, showing denser regions at the chamber bottom and lower density near the filament. Density gradients contribute to the buoyancy-induced flow structure.

**3.3. Chemical reactions and species distribution**

Having established the flow and thermal landscape of the reactor, we now examine the interplay of chemical kinetics and species transport that governs deposition chemistry. Figure 5 presents the steady-state spatial distributions of key gas-phase species at a filament temperature of 1100 °C—selected to reflect an industrially relevant regime where both reaction and transport processes are active.

Figures 5a and 5b show that $SiCl_4$ and $H_2$ mole fractions are highest near the inlet and decline along the reactor length due to progressive consumption by gas-phase reactions. Notably, both species exhibit lower concentrations in the upper half of the reactor, where elevated temperatures—driven by buoyancy-enhanced flow—favor reaction R2 ($SiCl_4 + H_2 \rightarrow SiHCl_3 + HCl$). Depletion is particularly pronounced near the filament, where local heating further accelerates this reaction. $H_2$ also participates in surface reaction R4, contributing to its steeper decline relative to $SiCl_4$.

Quantitatively, at 1100 °C, 30.5% of $SiCl_4$ is converted to $SiHCl_3$ via R2, and 18.6% is consumed overall. The corresponding $H_2$ consumption is 27.0%. These values underscore the activation of reaction pathways under filament heating and validate the thermokinetic consistency of the model.

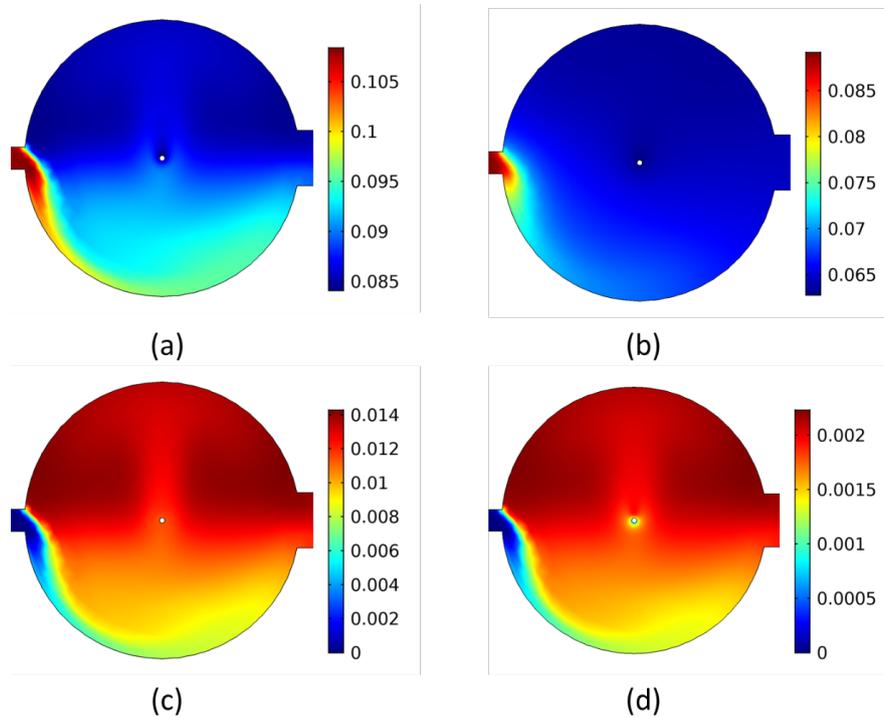

Figure 5. Steady-state mole fraction distributions of key species within the filament CVD reactor at a filament temperature of 1100 °C. (a) $SiCl_4$ mole fraction, showing progressive depletion along the flow path and localized consumption near the filament due to gas-phase reaction R2. (b) $H_2$ mole fraction, depleted by both gas-phase and surface reactions (R2 and R4), with concentrations highest in the lower chamber. (c) $SiHCl_3$ distribution, formed via R2 in intermediate temperature zones and depleted via R3 in the high-temperature region near the filament. (d) $SiCl_2$ mole fraction, produced through thermal decomposition of $SiHCl_3$ (R3) near the filament and consumed by surface deposition (R4). Species distributions reveal the temperature-sensitive sequence of reactions and their spatial coupling to flow and heat transfer.

The spatial profile of $SiHCl_3$ (Figure 5c) reflects its formation through R2 and subsequent depletion via R3 ($SiHCl_3 \rightarrow SiCl_2 + HCl$). $SiHCl_3$ concentrations rise as gas moves into thermally activated zones, peaking in the mid-to-upper regions of the chamber. However, in the highest temperature zone—immediately adjacent to the filament—$SiHCl_3$ levels drop due to the onset of R3. This spatial nonuniformity illustrates the sequential activation of reactions with increasing temperature and highlights the importance of understanding relative activation barriers: R3 has a higher activation energy than R2 (Table 1), so its influence only becomes significant at filament-level temperatures.

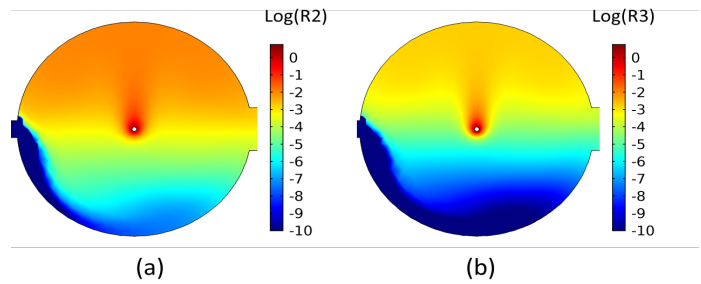

(a)             (b)

Figure 6. Logarithm of spatial distribution of gas-phase reaction rates at 1100 °C.(a) Reaction R2 ($SiCl_4 + H_2 \rightarrow SiHCl_3 + HCl$) is most active in the upper chamber, where buoyancy-driven heating raises local temperature.(b) Reaction R3 ($SiHCl_3 \rightarrow SiCl_2 + HCl$) is confined to the high-temperature region near the filament, consistent with its higher activation energy. These maps illustrate the spatial partitioning of sequential reactions and the effect of thermal gradients on reaction activation.

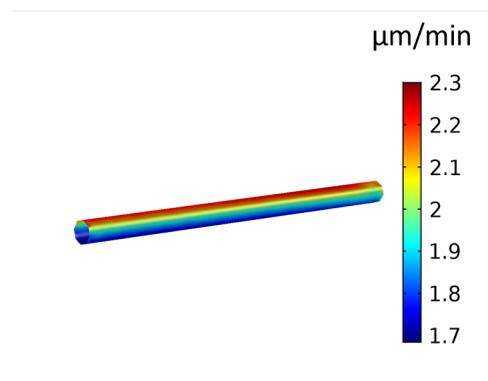

Figure 7. Computed Si deposition rate on the filament surface at 1100 °C, as governed by surface reaction R4. Growth rates range from 1.7 to 2.3 µm/min, with higher deposition observed on the upper filament surface due to local enrichment in reactive species and higher temperature. The non-uniform profile reflects the interplay of gas-phase reactions, transport phenomena, and surface kinetics, highlighting the importance of spatial resolution in process modeling.

$SiCl_2$ (Figure 5d), the terminal gas-phase precursor for surface deposition, exhibits a similar rise in the upper chamber and subsequent drop near the filament. This spatial signature reflects its dual role as a product of R3 and a reactant in the surface growth reaction R4. The zone of elevated $SiCl_2$ is tightly coupled to local reaction rates and gas circulation, reinforcing the need for spatially resolved modeling.

To further dissect these dynamics, Figure 6 presents the domain-wide distributions of gas-phase reaction rates for R2 and R3 (log scale), again at 1100 °C. Reaction R2 dominates in the upper chamber where moderate temperatures prevail, while R3 is strongly localized around the filament where temperatures are highest. The minimal R3 activity away from the filament supports the observed build-up of $SiHCl_3$ in intermediate zones and confirms that the onset of $SiCl_2$ production is thermally gated.

These spatial patterns are directly relevant for deposition design. The surface growth rate, computed via equation (15) and shown in Figure 7, ranges from 1.7 to 2.3 μm/min along the filament. A distinct asymmetry is observed, with higher growth localized at the top of the filament—corresponding to the region where thermally driven flow concentrates reactive species. This profile is not uniform and would be challenging to infer experimentally without intrusive diagnostics, further underscoring the value of high-fidelity CFD modeling.

Taken together, these results demonstrate how localized flow, heat transfer, and reaction kinetics interact to shape species distributions and deposition behavior. They also highlight the utility of computational models for diagnosing regime transitions, optimizing precursor utilization, and guiding filament placement and operating conditions in future reactor designs.

**3.4. Effect of process conditions**

Temperature is often treated as a lever to "ramp up" CVD chemistry. However, this simplification risks obscuring the nuanced ways in which temperature not only activates reactions but reshapes the dominant rate-limiting mechanisms. Here, we use the validated CFD framework to interrogate how temperature modulates deposition rate, growth uniformity, and precursor utilization across a wide process window (900–1700 °C), with the goal of refining design logic beyond the standard Arrhenius perspective.

3.4.1 Deposition Regimes and Rate Behavior

Figure 8 presents the Arrhenius plot of Si film growth rate on the filament surface. At lower temperatures (900–1100 °C), the process is reaction-limited. Growth rates increase exponentially with temperature, consistent with the activation energies of the dominant mechanisms: 73.7 kcal/mol for the thermal decomposition of $SiHCl_3$ (R3) and 67 kcal/mol for surface deposition via $SiCl_2$ and $H_2$ (R4). This exponential region reflects the thermally activated kinetics of both gas-phase and surface reactions.

In the intermediate temperature range (1100–1500 °C), the system enters a transitional regime. Reaction rates plateau, indicating full activation of chemical steps. Growth becomes increasingly constrained by mass transfer and precursor distribution rather than intrinsic reaction kinetics. The maximum growth rate—11.36 μm/min—is reached at 1400 °C.

Surprisingly, further temperature increases lead to a decline in growth rate. At 1500–1700 °C, buoyancy-driven flows accelerate gas movement, reducing residence time in high-temperature zones. As a result, reactive species are swept through the chamber before fully participating in deposition chemistry. This is the diffusion-limited regime, where transport—not chemistry—becomes the bottleneck. These observations challenge the implicit assumption that higher temperature always enhances film growth, underscoring the need to account for thermo-fluidic consequences in reactor design.

Our results are in strong agreement with literature data on Si deposition from chlorosilanes [12, 13, 15, 17, 33, 51], which consistently report maximal growth rates above 1100 °C. However, unlike most experimental studies, this model explicitly resolves how competing mechanisms shape the three canonical regimes: reaction-limited, transition, and diffusion-limited.

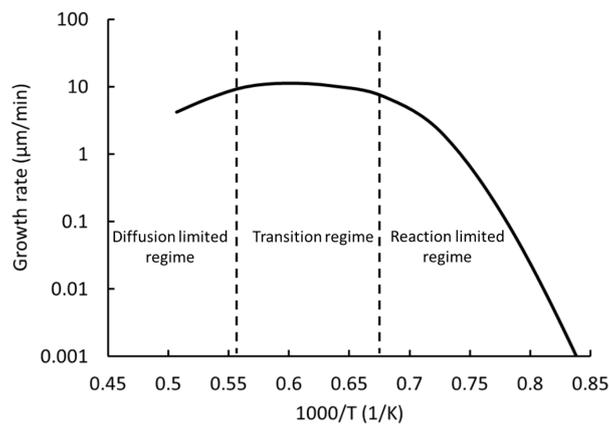

Figure 8. Arrhenius plot of Si film growth rate as a function of inverse filament temperature (900–1700 °C), illustrating the transition between reaction-limited, kinetically-saturated, and diffusion-limited regimes. The exponential increase at lower temperatures reflects activation of gas-phase and surface reactions, while the plateau and subsequent decline at higher temperatures indicate transport constraints and reduced residence time.

**3.4.2 Temperature Effects on Uniformity**

In many applications, deposition rate alone is not the performance target—uniformity is equally critical. To quantify this, we define the maximum surface non-uniformity as:

Maximum Non-uniformity = (max(GR) - min(GR)) / (average(GR)) × 100

Figure 9 plots this metric across the temperature range. At low temperatures (e.g., 900 °C), non-uniformity is high (~21%) due to partial activation of surface reactions, which favors deposition at hotter regions of the filament. As temperature rises to 1000–1100 °C, reactions become more uniformly active, and the gas-phase species distribution flattens, reducing non-uniformity (~17.5%).

However, this trend reverses at higher temperatures. Once in the diffusion-limited regime, transport gradients—not kinetics—dictate spatial variation. Specifically, precursor availability becomes asymmetric, and uniformity degrades again.

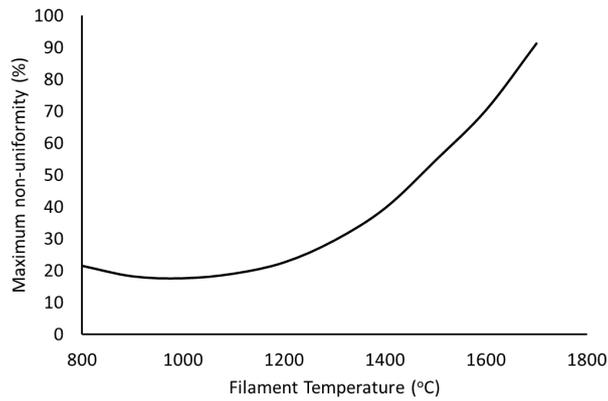

Figure 9. Maximum deposition non-uniformity across the filament surface as a function of filament temperature. The plot highlights a non-monotonic trend: uniformity improves in the reaction-limited regime due to more homogeneous activation, but degrades at higher temperatures as the process enters a diffusion-limited regime dominated by precursor transport asymmetries.

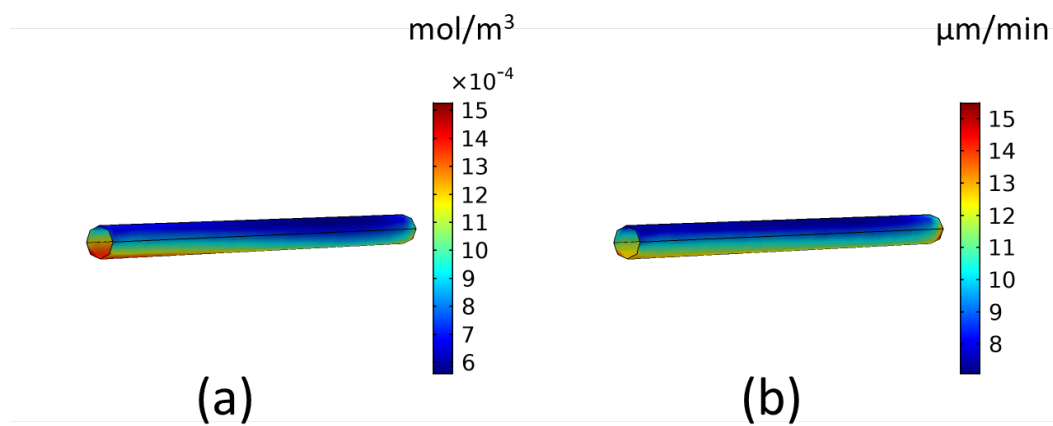

Figure 10. Spatial distribution of (a) hydrogen ($H_2$) concentration and (b) silicon film growth rate along the filament surface at a filament temperature of 1500 °C. The correlation between the two profiles reveals that in the diffusion-limited regime, local $H_2$ availability governs deposition rate, with higher growth occurring on the lower filament surface where $H_2$ concentration remains elevated.

A deeper mechanistic explanation is revealed by comparing the growth rate and $H_2$ concentration along the filament surface at 1500 °C (Figure 10). The observed trend is rooted in the stoichiometry of the overall reaction (R1), which requires an ideal $H_2$/$SiCl_4$ ratio of 2:1 for complete silicon dechlorination. In the current setup, $SiCl_4$ is introduced at 40 sccm, while $H_2$ enters at only 33 sccm, creating a potential local hydrogen deficit. As shown previously (Figure 5), hydrogen is preferentially depleted in the hotter upper region of the reactor, resulting in a vertical gradient where $H_2$ availability is significantly higher near the bottom.

This asymmetry in H₂ distribution becomes the dominant factor shaping the deposition profile at high temperatures. Figure 10 illustrates this clearly for the case of 1500 °C: the growth rate profile on the filament surface mirrors the local $H_2$ concentration, with higher deposition rates observed on the underside of the filament where $H_2$ remains more abundant. This is the inverse of the behavior observed at 1100 °C (Figure 7), where the top surface—closer to the hotter zone—dominates growth. These findings underscore the importance of considering both global stoichiometry and local precursor gradients when optimizing for deposition uniformity—especially in regimes where transport limitations dominate.

### 3.4.3 Implications for Design and Operation

The conventional view of temperature as a scalar control parameter must be revised. In filament CVD, temperature fundamentally reorganizes the deposition regime by reshaping species distributions, reaction balances, and transport dynamics. This analysis exposes how optimal temperature is not simply a balance between activation and thermal stability—but a *negotiation* between kinetics and flow-induced asymmetries.

The transition from reaction- to diffusion-limited behavior brings with it a shift in design priorities: from activating chemistry to engineering flow, precursor ratios, and filament placement to sustain uniformity and efficient utilization. Both the growth rate and uniformity are directly affected by the process regime, which in turn is determined by the limiting mechanism: thermal activation or precursor availability and transport. This decoupling between the effects of thermal activation and precursor availability underscores a key insight: in diffusion-limited regimes, uniformity and rate are no longer controlled by reactor temperature alone but by the local balance of species transport and consumption. These types of nuanced, geometry- and flow-sensitive behaviors are extremely difficult to observe experimentally but emerge clearly in high-resolution CFD simulations.

More broadly, this section illustrates how CFD can move beyond "prediction" toward explanation, providing mechanistic clarity that not only validates observed trends but challenges design intuition and refines process logic. These findings underscore the importance of considering both global stoichiometry and local precursor gradients when optimizing for deposition uniformity—especially in regimes where transport limitations dominate. These shifts are not intuitive and would be nearly impossible to measure directly. Only through fully coupled, spatially resolved simulation can these effects be observed and understood.

### 3.4.4 Effect of inlet gas flows

After the results regarding the effect of temperature, the effect of the reactant inlet flow is investigated. The effect of the inlet flow rate of the two reactants on the Si film growth rate is presented below, in Figure 11, for a filament temperature of 1100°C and 1500°C. These two temperature values are selected as representative cases for the reaction limited and diffusion limited regimes, respectively.

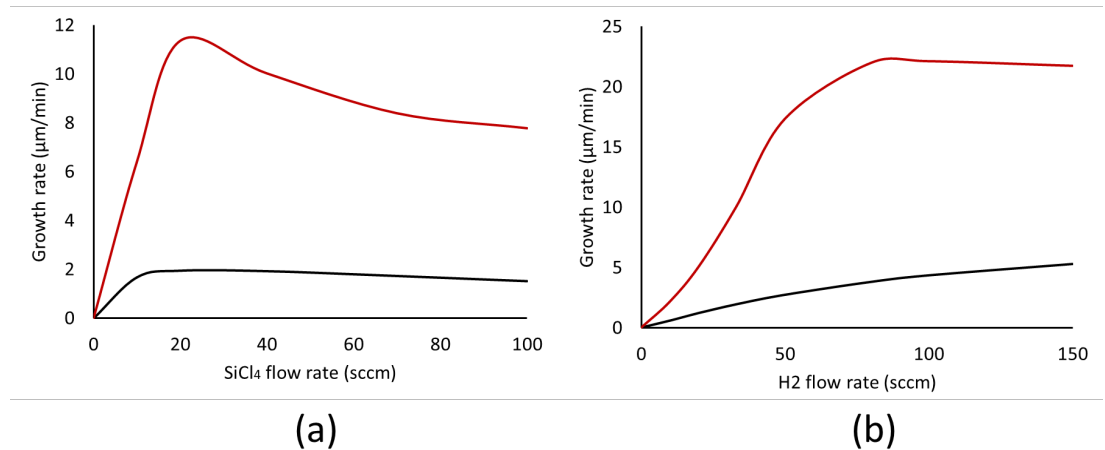

(a)                    (b)

Figure 11. Growth rate as a function of (a) $SiCl_4$ inflow rate and (b) $H_2$ inflow rate, evaluated at filament temperatures of 1100 °C (black) and 1500 °C (red). The comparison illustrates how temperature influences the sensitivity of deposition rate to precursor flow conditions.

Figure 11a shows that for low $SiCl_4$ flow rates (0-20sccm), an increase in the $SiCl_4$ flow rate leads to an increase of the growth rate, as expected for reactant flows. However, further increase of the $SiCl_4$ flow rate leads to a gradual decrease of the film growth rate, resulting in the complex behavior shown in Figure 11a. This decrease is attributed to the fact that, above 20 sccm, $SiCl_4$ becomes the abundant species for the overall reaction in the process (reaction R1). Therefore, for a $SiCl_4$ flow rate greater than 20 sccm, it is the $H_2$ reactant that limits the process. $SiCl_4$ flow increase beyond that point only leads to higher gas velocity, reducing the residence time of reactants in the CVD chamber, hence leading to less reactants depositing on the filament surface. This effect becomes more significant at higher temperature, as at 1500°C, the process is within the diffusion limited regime, and the reactions are thermally activated. In this regime, it is the reactant flow that dictates the Si film growth, and flow rate variations lead to more significant effect on the growth rate.

Figure 11b shows a significant effect of the $H_2$ flow on the growth rate. For the filament temperature of 1100°C, the process is still in the reaction limited regime, and the growth rate increases with the increase of the $H_2$ flow rate. Even when $H_2$ is in theoretical abundance (for $H_2$ flow rates above 80 sccm), the reaction is not fully activated, and not all of the reactants are reacting to deposit Si. Therefore, within the range of the tested $H_2$ flow rates (0-150 sccm), the Si film growth rate increases with the $H_2$ inflow.

On the contrary, for the filament temperature of 1500°C, the process is within the diffusion limited regime, and the reactions are fully activated. In this regime, the Si film growth rate increases with the $H_2$ flow rate, up to 80 sccm of $H_2$, which is the theoretical balance for the overall reaction (reaction R1) between the SiCl$_4$ (40 sccm) and H$_2$ flow rates. Further increase of the H$_2$ flow rate leads to a slight decrease of the growth rate, as the gas velocity increase starts to influence the film growth, as the residence time of the gas reactants within the CVD chamber decreases.

This different behavior of the film growth as a function of the reactants flow rate for the different filament temperature can be mechanistically explained by the model, enhancing the understanding of the process.

### 3.5. Effect of filament thickness

Following the study of the effect of the different process parameters on the film growth rate and the different aspects of the process, the analysis proceeds with the study of the effect of the filament design, and particularly the filament thickness. For this reason, the process was simulated using a filament diameter of 500μm. To illustrate the effect of the filament thickness on the Si film deposition, the Arrhenius plot for the 500μm filament and the 2mm filament are presented in Figure 12.

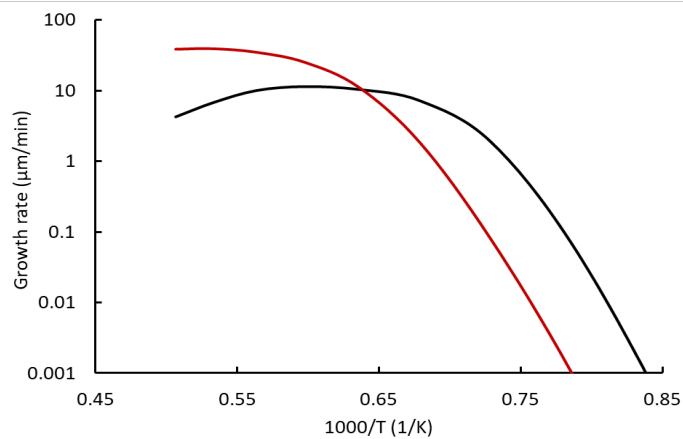

Figure 12. Arrhenius plot of silicon film growth rate as a function of inverse filament temperature for two filament diameters: 2 mm (black) and 500 μm (red). The plot highlights the impact of filament thickness on the thermal environment and deposition kinetics. At lower temperatures, the thicker filament enables higher growth rates due to more effective heating and gas-phase activation. At elevated temperatures, however, the thinner filament achieves superior deposition performance, attributed to reduced convective losses and enhanced local retention of reactive species. This transition underscores the importance of filament geometry in determining rate-limiting mechanisms across temperature regimes.

Figure 12 reveals the critical influence of filament diameter on Si film growth dynamics. At temperatures below 1100 °C, thinner filaments (500 μm) yield significantly lower deposition rates compared to their thicker counterparts (2 mm). This difference arises from the reduced radiative and convective heating capacity of the thinner filament, which results in lower gas-phase temperatures away from the filament surface. Consequently, the activation of gas-phase reactions—particularly $SiCl_4$ conversion and subsequent $SiHCl_3$ decomposition—is severely limited, restricting the formation of reactive intermediates and ultimately suppressing film growth.

As temperature increases, both filament configurations exhibit a similar exponential rise in growth rate, indicative of comparable activation energies governing the reaction-limited regime. However, beyond ~1300 °C, a reversal emerges: the 500 μm filament achieves higher growth rates than the 2 mm one. This counterintuitive behavior stems from the hydrodynamic and thermal characteristics of the smaller filament. At elevated temperatures, where reactions are fully activated, the thinner filament creates a tighter boundary layer and lower local gas velocities, which reduce precursor depletion via convective transport. The result is enhanced retention of reactive species near the surface and more efficient deposition. The growth rate peaks at ~39 μm/min around 1600 °C before slightly declining, marking the onset of a diffusion-limited regime where reactant availability becomes the dominant constraint.

These results underscore a fundamental design trade-off: thinner filaments can unlock higher deposition rates, but only within temperature regimes where full activation of surface and gas-phase chemistry is assured. Since filament geometry is often dictated by downstream application requirements (e.g., mechanical or optical constraints), this analysis highlights the necessity of carefully tailoring process conditions to the specific thermal and transport profiles induced by filament design.

**3.6. Multiple filament setup**

Building on the preceding analysis of filament thickness, we extend the investigation to multi-filament configurations—an essential step toward scaling up filament CVD processes for industrial relevance. To probe the interaction between neighboring filaments and its implications for deposition uniformity and efficiency, we introduced a second filament into the reactor domain and evaluated two canonical arrangements: horizontal and vertical alignment. In both cases, filaments were symmetrically placed 3 cm apart—either ±1.5 cm along the vertical axis or ±1.5 cm along the horizontal axis relative to the reactor's centerline.

Figure 13 presents the resulting temperature fields for both dual-filament configurations at a nominal filament temperature of 1100 °C, alongside the baseline single-filament case for reference.

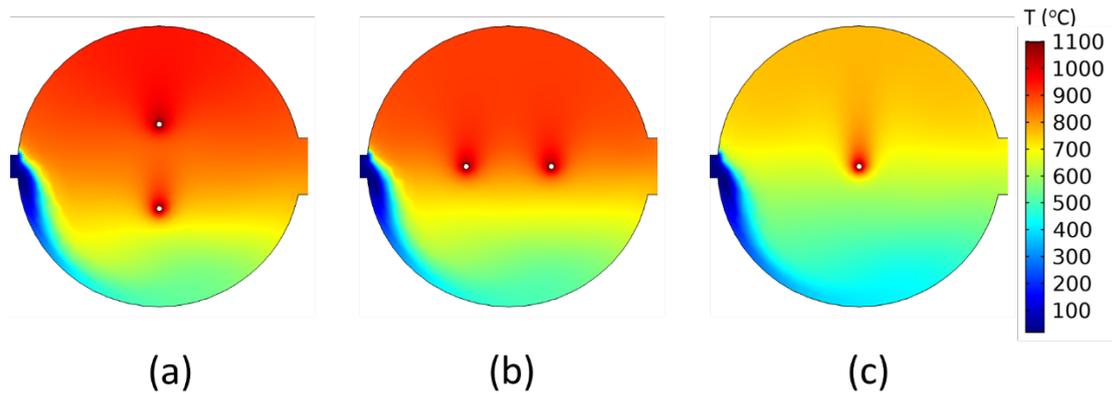

Figure 13. Temperature distribution within the reactor chamber for single-filament and dual-filament configurations (horizontal and vertical alignment), each operated at a filament temperature of 1100 °C. The addition of a second filament significantly broadens and elevates the heated region, particularly in the upper half of the chamber, due to increased total thermal input. These changes highlight the nonlinear impact of filament arrangement on reactor thermal fields and set the stage for altered gas-phase kinetics and transport behavior.

Figure 13 reveals a fundamental thermophysical consequence of introducing multiple filaments into the CVD reactor: a substantial elevation in the overall gas-phase temperature. Since the heated filaments are the primary energy source in this reactor configuration, increasing the number of filaments naturally increases the total radiative and convective heating surface. This results in a broader thermal envelope within the chamber. Specifically, in the dual-filament configurations, the upper half of the reactor reaches temperatures exceeding 900 °C—significantly higher than the ~750 °C observed in the single-filament case. To interrogate this further, we analyzed the influence of filament temperature on both average growth rate and maximum non-uniformity across the different filament configurations. The results, summarized in Figure 14, provide quantitative insight into how multi-filament setups reshape the process window—and potentially redefine optimal operating strategies.

The data presented in Figure 14a demonstrate a striking divergence in growth rate behavior between single and dual-filament configurations, particularly in the low-temperature regime. For filament temperatures below 1100 °C—where the process is clearly reaction-limited—the inclusion of an additional heated filament markedly elevates the average Si growth rate. While the single-filament setup yields a marginal growth rate (<0.1 μm/min) at 1000 °C, both multi-filament configurations achieve rates exceeding 2 μm/min under identical boundary conditions. This dramatic improvement is directly attributable to the spatial redistribution of thermal energy: multiple filaments effectively transform the chamber into a more uniformly heated environment, expanding the reaction zone volume and facilitating broader activation of gas-phase reactions (as corroborated by the thermal fields in Figure 13).

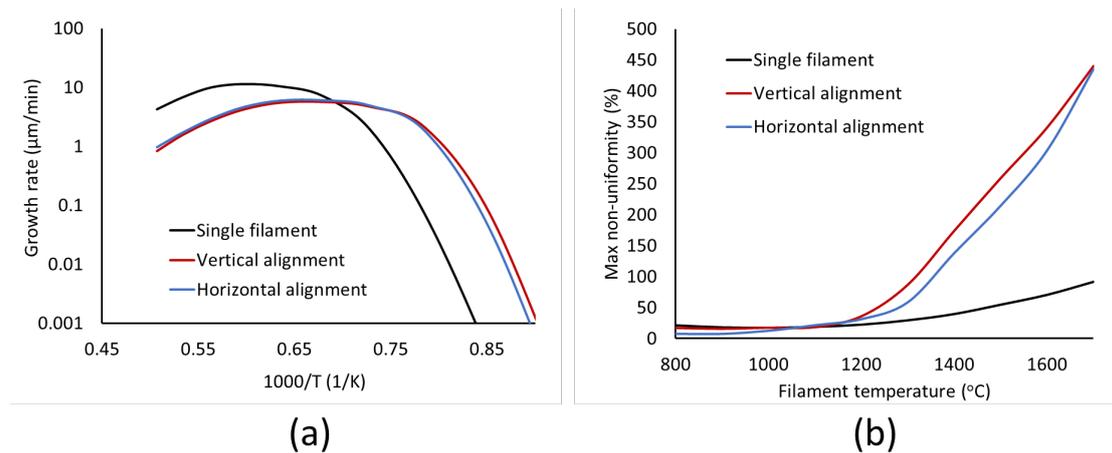

Figure 14. (a) Average Si deposition rate as a function of filament temperature for single-filament (black), dual-filament horizontal (red), and dual-filament vertical (blue) configurations. The results highlight the enhanced reactivity enabled by distributed heating in the reaction-limited regime (<1100 °C), and the transition to reactant-limited performance at higher temperatures. (b) Maximum non-uniformity of Si deposition as a function of filament temperature for single and dual-filament configurations. While multi-filament setups improve uniformity under reaction-limited conditions through thermal homogenization, they introduce significant spatial disparities in reactant availability—and thus higher non-uniformity—in the diffusion-limited regime.

This advantage, however, is transient. As the filament temperature increases beyond ~1200 °C and the system transitions into the reaction-saturated regime, the benefits of distributed heating begin to plateau. Although all configurations achieve their respective maximum growth rates in this zone, the single-filament setup outperforms its multi-filament counterparts, achieving ~11 μm/min compared to ~6 μm/min for the dual-filament cases. This inverse behavior stems from a fundamental constraint: total precursor throughput is constant across all scenarios. When deposition surfaces multiply, reactant species are divided among them. Hence, while distributed heating enhances early-stage kinetics, it dilutes peak deposition per unit area in the saturation regime.

As the system enters the diffusion-limited regime at temperatures exceeding ~1400 °C, this trade-off becomes more pronounced. The extended high-temperature zones created by multiple filaments induce stronger buoyancy and faster convective turnover, thereby decreasing gas-phase residence times. These transport-driven effects reduce effective precursor utilization and depress growth rates more steeply than in the single-filament case. The very mechanism that enabled growth at low temperatures—thermal homogenization—now accelerates depletion and disrupts surface saturation dynamics.

Uniformity trends, shown in Figure 14b, follow a similarly bifurcated logic. In the reaction-limited regime, dual-filament setups offer superior uniformity due to more even temperature fields and symmetric reaction zones. The horizontally aligned filaments perform especially well in this regard, benefiting from near-identical exposure profiles. In contrast, vertical alignment introduces an intrinsic asymmetry: the upper filament resides in a hotter, reaction-favored region, while the lower filament experiences cooler, kinetically suppressed conditions—leading to an intrinsic non-uniformity baked into the geometry.

This advantage, however, inverts beyond 1200 °C. In the diffusion-limited regime, gas-phase transport dominates deposition behavior, and reactant availability—particularly of $H_2$—becomes the critical bottleneck. As previously established, $H_2$ tends to accumulate in cooler lower regions and depletes rapidly in hotter upper zones. The more extensive thermal footprint of the dual-filament setups intensifies this stratification, particularly in the vertical configuration, where filaments are separated across these gradients. The resulting spatial disparity in $H_2$ concentration manifests directly as non-uniform deposition, with the vertical configuration exhibiting the highest deviation at elevated temperatures. Instead, this study shows that filament multiplicity introduces competing effects—enhancing growth under kinetically limited conditions but undermining both rate and uniformity in transport-limited regimes. Optimal design thus requires an integrated view of thermal management, flow structure, and chemical kinetics. Computational modeling proves indispensable here, offering insights that would be prohibitively difficult to extract from experimental campaigns alone.

**3.6. Sensitivity Analysis**

To understand how different process conditions influence silicon film deposition, we performed a sensitivity analysis using Polynomial Chaos Expansion (PCE), presented in Section 2.6. This method allows us to estimate how much each input parameter—filament temperature, and the inlet concentrations of $H_2$ and $SiCl_4$—affects the outcome, specifically the deposition rate of the silicon film.

Each of these parameters was treated as uncertain within its expected operating range, and we assumed they could vary evenly across that range (i.e., we used a uniform distribution). This simplification enables the use of a structured set of mathematical functions (Legendre polynomials) to build a fast, approximate version of our simulation model—known as a PCE surrogate model. The coefficients of this surrogate model were calculated using least squares regression. A brief convergence test showed that using polynomials up to degree 3 was sufficient to achieve good accuracy.

Once this surrogate model was built, we used it to perform a global sensitivity analysis. Specifically, we calculated Sobol' indices, which tell us how much each input parameter (or combination of parameters) contributes to variability in the predicted deposition rate. The results are shown in Figure 15 and are separated into two key temperature regimes:

(i) Reaction-limited regime (800–1100 °C): In this range, the reaction rates are relatively slow, and the process is limited by how fast chemical reactions occur. As shown in Figure 15a, filament temperature and $SiCl_4$ concentration have the strongest influence on deposition. This reflects the fact that reactions depend heavily on thermal energy and precursor availability in this regime; (ii) Diffusion-limited regime (>1500 °C): At these higher temperatures, the chemical reactions are already fast, so the process becomes limited by how quickly reactants can reach the filament surface. In this case (Figure 15c), temperature still matters, but for a different reason: it now affects gas flow and decomposition, which can lead to depletion of available precursors. Meanwhile, the influence of $SiCl_4$ and $H_2$ concentrations becomes more prominent, since they control how much reactant is available when transport becomes the bottleneck.

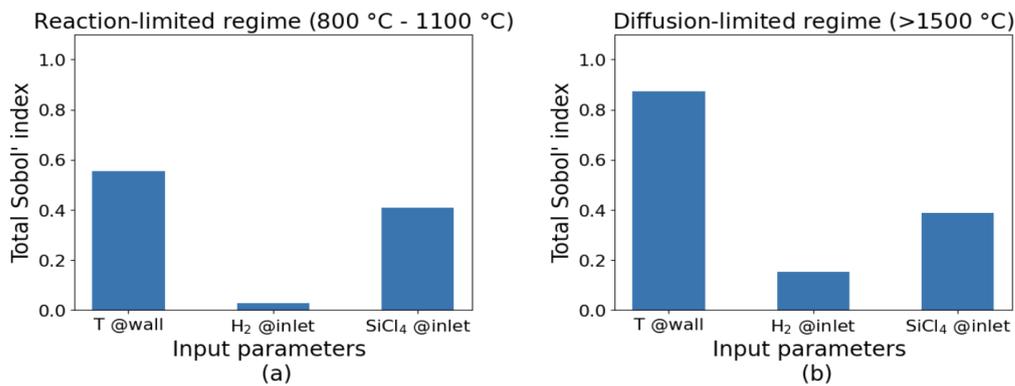

Figure 15. Total Sobol' indices quantifying the influence of each input parameter—filament temperature, and inlet concentrations of $H_2$ and $SiCl_4$—on the variance of the predicted Si deposition rate across different regimes of the Arrhenius plot. Each index accounts for both the direct effect of a parameter and its interactions with others, providing a comprehensive measure of sensitivity in the reaction-limited and diffusion-limited regimes.

## 4. Conclusions

This study introduces a fully coupled, three-dimensional Computational Fluid Dynamics (CFD) model for silicon (Si) Chemical Vapor Deposition (CVD) on tungsten (W) filaments, incorporating all major physical and chemical mechanisms—from gas flow and heat transfer to gas-phase and surface reactions. Validated against three independent experimental datasets, the model delivers not only predictive power but also diagnostic clarity, offering a unified framework to understand and redesign filament-based CVD systems.

A central insight of this work is that the filament is not a passive deposition surface but an active thermal and flow-controlling element that fundamentally reshapes the reactor environment. Buoyancy-driven convection, localized temperature gradients, and non-uniform species distribution emerge as primary factors influencing deposition performance. These effects are non-trivial—and often overlooked—in simpler models based on planar geometries or 1D approximations.

By systematically exploring the process across temperatures, geometries, and configurations, the model delineates three key operating regimes: reaction-limited, transition, and diffusion-limited. Each regime is characterized by a different controlling mechanism, revealing how deposition efficiency and uniformity evolve with temperature and gas composition. Such mechanistic segmentation is critical for designing targeted process controls and operational windows.

Furthermore, the model identifies trade-offs between growth rate and uniformity as a function of filament thickness and arrangement. Notably, thinner filaments outperform thicker ones at high temperatures due to improved local confinement, while multiple filaments enhance deposition in low-temperature regimes but exacerbate non-uniformities under high-throughput conditions. These results underscore the importance of tailoring both geometry and thermal management to the specific deposition regime.

Global sensitivity analysis, using Polynomial Chaos Expansion and Sobol' indices, adds a quantitative layer to these findings. It reveals how the influence of control parameters—such as filament temperature and precursor flow rates—shifts between regimes, offering actionable guidance for robust process optimization.

Crucially, the insights gained from this study are not limited to lab-scale demonstrations. The modeling approach provides a foundation for rational scale-up of filament CVD systems. By diagnosing the interplay between flow, heat, and reaction kinetics, the model enables predictive control over reactor behavior at larger scales. This opens the door to industrial-scale deployment of filament-based CVD for high-throughput production of Si films—an approach that is potentially more energy-efficient, modular, and geometrically flexible than conventional planar CVD.

In sum, this work reframes filament CVD as a distinct technological paradigm—one that demands its own mechanistic understanding and design logic. By bridging deep physical modeling with systems-level insight, it offers a powerful roadmap for advancing both the science and engineering of scalable silicon deposition.

**Acknowledgements**

The current work has received funding from the Hellenic Foundation for Research and Innovation (HFRI) program Data4Solar, project ID No 16568. This research was funded in part by the Luxembourg National Research Fund (FNR), grant reference [16758846].